\documentclass[12pt]{article}

\usepackage{graphicx}
\usepackage{amsmath}

\begin{document}

\begin{center}

{\Large {\bf Wigner representation for polarization-momentum hyperentanglement
generated in parametric down-conversion, and its application to
complete Bell-state measurement}}

\end{center}

\vspace{0.5cm}

\begin{center}

{\bf A. Casado$^1$, S. Guerra$^2$, and J. Pl\'{a}cido$^3$}.

$^1$ Departamento de F\'{\i}sica Aplicada III, Escuela T\'{e}cnica Superior de
Ingenier\'{\i}a,

Universidad de Sevilla, 41092 Sevilla, Spain.

Electronic address: acasado@us.es

$^2$ Centro Asociado de la Universidad Nacional de Educaci\'on a
Distancia de Las Palmas de Gran Canaria,

 35004 Las Palmas de Gran Canaria, Spain.

$^3$ Grupo de Ingenier\'{\i}a T\'{e}rmica e Instrumentaci\'{o}n, Universidad de Las
Palmas de Gran Canaria,

 35017 Las Palmas de Gran Canaria, Spain.
\vspace{0.25cm}
\end{center}


\vspace{0.25cm}
\noindent {\bf Abstract}

We apply the Wigner function formalism to the study of two-photon
polarization-momentum hyperentanglement generated in parametric
down-conversion. It is shown that the consideration of a higher number
of degrees of freedom is directly related to the extraction of
additional uncorrelated sets of zeropoint modes at the source. We
present a general expression for the description of the quantum
correlations corresponding to the sixteen Bell base states, in terms of
four beams whose amplitudes are correlated through the stochastic
properties of the zeropoint field. A detailed analysis of the two
experiments on complete Bell-state measurement included in [Walborn et
al., Phys. Rev. A {\bf 68}, 042313 (2003)] is made, emphasizing the
role of the zeropoint field. Finally, we investigate the relationship
between the zeropoint inputs at the source and the analysers, and the
limits on optimal Bell-state measurement.


\vspace{0.5cm}

\section{INTRODUCTION}
In the last two decades, parametric down-conversion (PDC) assumed an
important role for the practical implementation of the quantum theory
of information, such as quantum cryptography \cite{qcrypt}, quantum
dense coding \cite{densecod} and teleportation \cite{telep}. The use of
PDC as a source of entanglement involves the necessity of performing a
complete Bell-state measurement (BSM), which is required in many
quantum communication schemes.  In this context, entangled photon pairs
produced in PDC have been used for experiments on a partial Bell-state
measurement \cite{mattle}, in which entanglement involves only one
degree of freedom, and a complete Bell-state measurement, in which
hyperentanglement (entanglement between two or more degrees of freedom)
takes part \cite{kwiatweinfurter}. More recently, these states have
been used as an essential part of cluster states \cite{Vallone} which are of
great value in the field of quantum computing \cite{Tame}.

The use of enlarged Hilbert spaces opened the door for a complete BSM
using linear optics and single photon detectors, first by considering
that one of the degrees of freedom was in a fixed quantum state
\cite{Walborn}, and encoding the information in the other. Further
studies about the actual limits for BSM in these enlarged spaces have
shown that, for two photons hyperentangled in $n$ degrees of freedom,
the number of mutually distinguishable sets of Bell states is bounded
above by $2^{n+1}$ \cite{2007}. More recently, it has been shown that
at most $2^{n+1}-1$ classes out of $4^n$ hyper-Bell states can be
distinguished with one copy of the input state, and that complete
distinguishability is possible with two copies, within the class of
devices obeying linear evolution and local measurement (LELM)
\cite{2011}. In case the two photons are not brought together at the
LELM apparatus the maximun number of distinguishable classes is $2^n$.

The Wigner representation of quantum optics provides an alternative to
the standard Hilbert space formalism for the study of quantum
information and for its practical implementation with PDC. In the
Wigner representation within the Heisenberg picture (WRHP) the
generation and propagation of PDC light is treated as in classical
optics by taking into account the zeropoint field (ZPF) entering the
crystal and the different optical devices placed between the source and
the detectors. Finally, the vacuum fluctuations of the electromagnetic
field are subtracted at the detectors \cite{pdc2, pdc4}. Hence, the
peculiarities of the quantum world with respect to the image that
classical physics offers are represented, in this context, by (i) the
existence of a stochastic zeropoint field whose amplitudes are
distributed according to a positive Wigner function, and (ii) the way
in which the signal is separated from the zeropoint background in the
detection process. These two features give rise to the typical
counterintuitive results within the quantum domain.

In essence, manipulating entanglement is a common denominator in the
different manifestations of quantum communication. In the WRHP,
two-photon polarization entanglement is represented by two stochastic
light beams, whose correlation properties arise from the coupling
between two zeropoint beams (each containing two sets of uncorrelated
zeropoint modes) and the laser beam at the crystal \cite{pdc4},
following the classical Maxwell equations \cite{pdc5, pdc6}. On the
other hand, entanglement manipulation involves the change in the
correlation properties of the light beams when they are going through
the different optical devices, and this is related to the way in which
the vacuum modes are redistributed at the field amplitudes.

The WRHP description of the four polarization Bell base states was made
in \cite{pdc7} along with the application of the formalism to
experiments on quantum cryptography. In \cite{pdc8} partial Bell-state
analysis was studied, and the fermionic behaviour of two photons
described by the singlet state, when they reach a balanced
beam-splitter, was explained using purely wave-mechanical arguments
based on the Wigner representation. More recently, the WRHP formalism
has been applied to the description of entanglement swapping using PDC
light, and it has been shown that the generation of mode entanglement
between two initially non interacting photons is related to the
quadruple correlation properties of the electromagnetic field, through
the stochastic properties of the vacuum \cite{swapping}. These works
emphasised the role of the zeropoint field in the generation and
propagation of light in experimental implementations of quantum
communication, including the existence of a relevant ZPF noise entering
the idle channels of the analysers. Concretely, according to
\cite{pdc7} the effects of eavesdropping attacks in the case of
projective measurements are directly related to the inclusion of some
fundamental noise, that also turns out to be fundamental to reproduce
the quantum results. In this way, we can state that the zeropoint field
carries the quantum information which is extracted at the source, and
also introduces some fundamental noise at the idle channels of the
analyzers. These two features of the zeropoint field are a common
denominator in optical experiments on quantum information.

The standard Hilbert-space formulation of quantum optics considers
vacuum fluctuations in an implicit way through the Heisenberg principle
and the use of normal ordering for the calculation of photodetection
probabilities. In contrast, the Wigner function offers the possibility
of stating specifically what the role of vacuum fluctuations is in the
generation and measurement of quantum information in quantum optical
information processing using PDC. In this way, the motivation for this
paper and further works comes from the following questions: What is the
relationship between enlarging the Hilbert space and the zeropoint
field activated at the source? What is the role of the zeropoint at the
different stages of a BSM experiment? What is the relationship between
the zeropoint modes entering the source, the ones activated at the idle
channels of the analyzers, and the maximal information that can be
generated in each experiment? Thus, there is considerable motivation
for the application of the WRHP approach to the description of
hyperentanglement and its application to complete Bell-state analysis,
in order to investigate the role of the zeropoint field in this area.

The paper is organised as follows: In Section \ref{sec2} we shall
describe two-photon polarization-momentum hyperentanglement generated
in PDC, by means of four correlated light beams. This description
involves the consideration of eight uncorrelated sets of vacuum modes,
distributed in four ZPF entering beams at the source. We shall obtain a
compact expression for the description of the sixteen Bell base states
in the WRHP, in terms of four two-by-two correlated beams. In Section
\ref{sec3} we shall apply this formalism to study the experiments
proposed in reference \cite{Walborn} in which one of the degrees of
freedom is in a fixed quantum state. We shall put the emphasis on the
role of the zeropoint field at the different steps of each experiment,
in order to make clear its relevant contribution to the signal fields
arriving at the detectors. This analysis is important, not only for the
theoretical aspects concerning optical experiments on quantum
information, but also for its relation with optical tests of Bell's
inequalities, in which the vacuum field entering the idle channels of
the analysers gives rise to enhancement \cite{santos, david}. In
Section \ref{sec4} we shall demonstrate that the number of independent
sets of zeropoint modes entering the source represents an upper bound
to the maximum number of Bell states that can be distinguished in
hyperentanglement-assisted Bell-state analysis. Also, we shall
establish the relationship between the ZPF inputs at the source and the
analysers, and the maximum number of distinguishable Bell-state
classes, in LELM apparatus in which the left and right input channels
are not brought together. Finally, in Section \ref{sec5} we shall
present the main conclusions of this work, and sketch further steps for
future research. In order to a better understanding of the WRHP
approach in this paper, we have included some fundamental ideas in
Appendix \ref{Appendix}.

\section{POLARIZATION-MOMENTUM HYPERENTANGLEMENT IN THE WRHP}
\label{sec2}

Let us start by considering the following situation: a type-I
two-crystal source is pumped by a laser beam. The first (second)
crystal emits pairs of horizontal (vertical) polarized photons in
superimposed emission cones. Because the photons are emitted on
opposite sides of the cone, two sets of conjugated beams, $(a_1, b_2)$
and $(a_2, b_1)$, which are represented by wave vectors ${\bf
k}_{a_i}$, ${\bf k}_{b_i}$ ($i=1, 2)$, can be selected \cite{kwiat}. If
the coherence volume of the laser contains the two-crystal interaction
region, the quantum state corresponding to a photon pair is usually
expressed, in a particle-like description, as:

\begin{equation}
|\Phi^+\rangle \otimes |\psi^+\rangle=
\frac{1}{\sqrt{2}}[|H\rangle_1|H\rangle_2+
|V\rangle_1|V\rangle_2] \otimes
\frac{1}{\sqrt{2}}[|a\rangle_1|b\rangle_2+ |b\rangle_1|a\rangle_2].
\label{hyper6}
\end{equation}

The state given by (\ref{hyper6}) is one of the sixteen base states
corresponding to the two-photon hyperentanglement on polarization and
momentum degrees of freedom, of the form $| \Pi\rangle \otimes |
\eta \rangle$, where $| \Pi\rangle$ ($| \eta\rangle$) is the four-dimensional
vector representing one of the polarization (momentum) Bell base states
\cite{Walborn}:

\begin{equation}{\left| \Psi ^{\pm}  \right\rangle}  =\frac{1}{\sqrt{2} }
\left[{\left| H \right\rangle} _{1} {\left| V \right\rangle}_{2}
\pm {\left| V \right\rangle} _{1} {\left| H \right\rangle} _{2} \right]
\,\,\,;\,\,\,{\left| \Phi ^{\pm }  \right\rangle} =\frac{1}{\sqrt{2} }
\left[{\left| H \right\rangle} _{1} {\left| H \right\rangle} _{2}
\pm {\left| V \right\rangle} _{1} {\left| V \right\rangle} _{2} \right]
,\label{s1}\end{equation}

\begin{equation}
{\left| \psi ^{\pm }  \right\rangle} = \frac{1}{\sqrt{2} }
\left[{\left| a \right\rangle} _{1} {\left| b \right\rangle} _{2} \pm {\left| b \right\rangle} _{1}
{\left| a \right\rangle} _{2} \right]\,\,\,;\,\,\,{\left| \phi ^{\pm}
\right\rangle}= \frac{1}{\sqrt{2} }
\left[{\left| a \right\rangle} _{1} {\left| a \right\rangle} _{2} \pm {\left| b \right\rangle} _{1}
{\left| b \right\rangle} _{2} \right].
\label{s3}\end{equation}

\begin{figure}[h]
      \centering
      \includegraphics[height=6cm]{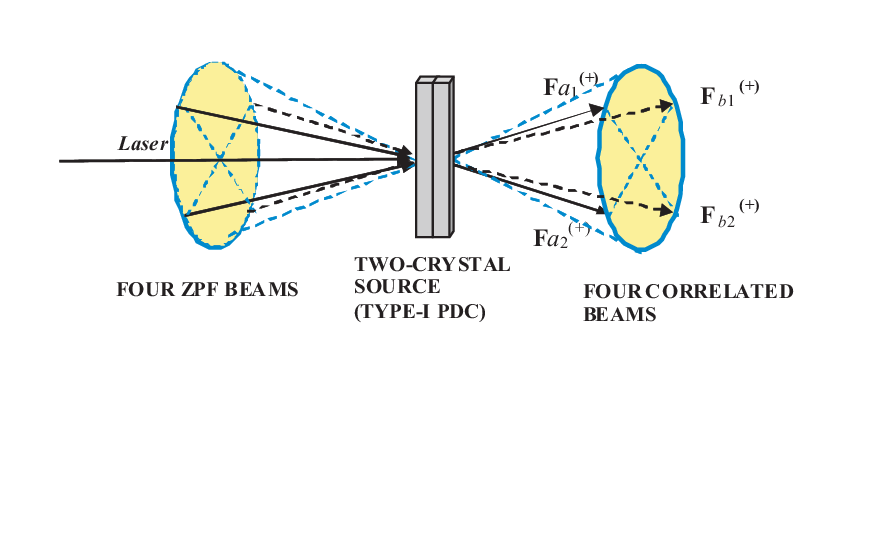}
\caption{
\small{Sets of vacuum modes (on the left), which are ``activated" and coupled with
the laser inside the crystal. The correlation properties of the four
beams (on the right) are related to the way in which the vacuum
amplitudes are distributed in the field amplitudes. Hence, each of the
sixteen Bell base states is characterized, in the WRHP approach, by a
``particular" set of correlations.} }
\label{fig1}
\end{figure}

The study of polarization-momentum hyperentanglement in the WRHP is
based on the same ideas that were developed in references \cite{pdc2}
and \cite{pdc4}. The key point in this case is that the selection of
two sets of correlated beams, $(a_1, b_2)$ and $(a_2, b_1)$, implies
the consideration of eight sets of vacuum modes which are ``activated"
at the crystal via the coupling with the laser beam (see
fig.\ref{fig1}). The set of representative modes, corresponding to the
entering zeropoint beam of wave vector ${\bf k}_{x_i}$, is represented
by the vacuum amplitudes:

\begin{equation}
\left\{\alpha_{x_{i}, \lambda} \right\} \equiv
\left\{\alpha _{{\bf k},\lambda};
{\bf k}\in \left[{\bf k}\right]_{x_{i}} \right\}\,\,\,;\,\,\,x=a, b
\,\,\,;\,\,\,\lambda=H,V\,\,;\,\,i=1,2.
\label{mode1}
\end{equation}



In order to focus on the main points we shall first describe the
generation of the beams concerning one of the sixteen states, the one
corresponding to Eq. (\ref{hyper6}). The Hamiltonian corresponding to
the electromagnetic field can be expressed in the following way:

\[
H=H_{free}+H_{int} =
\sum_{\lambda=H,V}
\sum_{\bf k}\hbar \omega_{{\bf k},\lambda}
\alpha_{{\bf k},\lambda}^{*}\alpha_{{\bf k},\lambda}
\]
\begin{equation}{+\left(i\hbar g'\frac{V}{2} \sum _{\begin{array}{l} {i,j=1}
\\ {i\ne j} \end{array}}^{2}{\kern 1pt} \sum _{\lambda =H,V}
\sum _{\begin{array}{l} {{\bf{k}}\in \left[{\bf{{\rm {\bf{k}}}}}\right]_{a_{i} } }
\\ {{\bf{k}}'\in \left[{\bf{{\rm {\bf{k}}}}}\right]_{b_{j} } }
\end{array}}
f\left({\bf{k}},{\bf{k'}}\right)\exp \left(-i\omega _{p} t\right)
\alpha _{{\bf{k}},\lambda }^{*} \alpha _{{\bf{k'}},\lambda }^{*} +\rm c.c.   \right),}
\label{free1}\end{equation}
where the crucial difference between (\ref{free1}) and Eq. (1) of
\cite{pdc4} is that we have made the change $V\rightarrow V/2$, in
order to consider that the energy of the classical wave corresponding
to the laser (with frequency $\omega_p$ and momentum ${\bf k}_p$),
which is proportional to the squared amplitude, must be divided into four
beams. On the other hand, $f\left({\bf{k}},{\bf{k'}}\right)$ is a
function which is different from zero only when the momentum matching
condition is fulfilled, and $g'$ is a constant related to the coupling
parameter.

The evolution equation for $\alpha_{{\bf k},
\lambda}$ is given by the Hamilton (canonical) equations, taking $\sqrt{\hbar}\alpha_{{\bf k},
\lambda}$ as coordinates and $\sqrt{\hbar}\alpha^*_{{\bf k},
\lambda}$ as canonical momenta. We have:

\begin{equation}
\dot{\alpha}_{{\bf k}, \lambda}=-i\omega_{{\bf k}, \lambda}
\alpha_{{\bf k}, \lambda}
+g'\frac{V}{2}\sum_{{\bf k}'}f({\bf k},{\bf k}') \exp(-i\omega_p t)
\alpha_{{\bf k}', \lambda}^{*},
\label{eqb5}
\end{equation}
where $\{{\bf k}, \lambda\}$ represents any mode belonging to
(\ref{mode1}). The integration is performed to second order in the
coupling constant ($g=g'\Delta t$), from $t=-\Delta t$ to $t=0$,
$\Delta t$ being the interaction time inside the two-crystal source.
For $t>0$ there is a free evolution.

By substituting $\alpha_{{\bf k}, H}(t)$ and $\alpha_{{\bf k}, V}(t)$
in Eq. (\ref{F}) we obtain the following four correlated beams leaving
the crystal (for more details see \cite{pdc2, pdc4}):

\begin{equation}
{\bf F}_{a_1}^{(+)}({\bf r}, t)=F_p^{(+)}({\bf r}, t; \{\alpha_{a_1, H}; \alpha^*_{b_2, H}\}){\bf
i}_{a_1}+F_{s}^{(+)}({\bf r}, t; \{\alpha_{a_1, V}; \alpha^*_{b_2,
V}\}){\bf j}_{a_1},
\label{hyper12}
\end{equation}

\begin{equation}
{\bf F}_{b_2}^{(+)}({\bf r}, t)=F_q^{(+)}({\bf r},
t; \{\alpha_{b_2, H}; \alpha^*_{a_1, H}\}){\bf
i}_{b_2}+F_{r}^{(+)}({\bf r}, t;  \{\alpha_{b_2, V}; \alpha^*_{a_1,
V}\}){\bf j}_{b_2},
\label{hyper13}
\end{equation}

\begin{equation}
{\bf F}_{b_1}^{(+)}({\bf r}, t)={F'}_p^{(+)}({\bf r},
t; \{\alpha_{b_1, H}; \alpha^*_{a_2, H}\}){\bf
i}_{b_1}+{F'}_{s}^{(+)}({\bf r}, t;  \{\alpha_{b_1, V}; \alpha^*_{a_2,
V}\}){\bf j}_{b_1},
\label{hyper15}
\end{equation}

\begin{equation}
{\bf F}_{a_2}^{(+)}({\bf r}, t)={F'}_q^{(+)}({\bf r},
t; \{\alpha_{a_2, H}; \alpha^*_{b_1, H}\}){\bf
i}_{a_2}+{F'}_{r}^{(+)}({\bf r}, t;  \{\alpha_{a_2, V}; \alpha^*_{b_1,
V}\}){\bf j}_{a_2},
\label{hyper14}
\end{equation}
where each polarization component is a linear transformation of the ZPF
entering the nonlinear medium. We have included the sets of relevant
zeropoint modes, for a better understanding of the correlation
properties. For instance, $F_p^{(+)}$ is only correlated to $F_q^{(+)}$
because $\{\alpha_{a_1, H}\}$ ($\{\alpha^*_{b_2, H}\}$) is correlated
to $\{\alpha^*_{a_1, H}\}$ ($\{\alpha_{b_2, H}\}$), as it can be seen
from Eq. (\ref{correlations}). The same reasoning applies to the
correlations $F_r^{(+)}
\leftrightarrow F_s^{(+)}$, ${F'}_p^{(+)}
\leftrightarrow {F'}_q^{(+)}$, and ${F'}_r^{(+)} \leftrightarrow
{F'}_s^{(+)}$. Hence, the beam ${\bf F}_{a_1}^{(+)}$ (${\bf
F}_{b_1}^{(+)}$) is correlated to ${\bf F}_{b_2}^{(+)}$ (${\bf
F}_{a_2}^{(+)}$), but there is no correlation between the two beams
corresponding to each photon.

Taking ${\bf r}={\bf 0}$ at the center of the source, we have the
cross-correlation:

\begin{equation}
\langle F_p^{(+)}({\bf 0}, t)F_q^{(+)}({\bf 0}, t')\rangle
=g\frac{V}{2}\nu (t'-t),
\label{hyper16}
\end{equation}
where $\nu (t'-t)$ vanishes when $|t'-t|$ is greater than the
correlation time between the beams ${\bf F}_{a_1}^{(+)}$ and ${\bf
F}_{b_2}^{(+)}$ \cite{pdc5}. Similar expressions hold for $\langle
{F'}_r^{(+)}({\bf 0}, t){F'}_s^{(+)}({\bf 0}, t')\rangle$, $\langle
F_r^{(+)}({\bf 0}, t)F_s^{(+)}({\bf 0}, t')\rangle$, and $\langle
{F'}_p^{(+)}({\bf 0}, t){F'}_q^{(+)}({\bf 0}, t')\rangle$.

On the other hand, taking for instance the polarization amplitude
$F_p^{(+)}$ at a point ${\bf r}$ and times $t$ and $t'$, we have the
autocorrelation:
\[
\langle F_p^{(+)}({\bf r},t)F_p^{(-)}({\bf r}, t') \rangle -\langle
[{\bf F}_{ZPF, a_1}^{(+)}({\bf r},t)\cdot
{\bf i}_{a_1}][{\bf F}_{ZPF, a_1}^{(-)}({\bf r},t')\cdot
{\bf i}_{a_1}]\rangle
\]
\begin{equation}
=\frac{g^2|V|^2}{4}\mu(t'-t),
\label{eqb}
\end{equation}
where ${\bf F}_{ZPF, a_1}^{(+)}$ is the zeropoint beam corresponding to
mode $a_1$, and $\mu(t-t')$ is a correlation function which goes to
zero when $|t'-t|$ is greater than the coherence time of PDC light.
Similar expressions hold for $F_s^{(+)}$, $F_q^{(+)}$, $F_r^{(+)}$, and
for the corresponding primed amplitudes.

\subsection{The sixteen hyper-Bell states in the WRHP}
The four beams given in Eqs. (\ref{hyper12}) to (\ref{hyper14}) are
correlated through the ZPF entering the two-crystal source, which is
``amplified" via the activation of the eight sets of vacuum modes
$\{{\bf k}_{x_i, \lambda}\,\,(x=a, b; i=1, 2; \lambda=H, V)\}$. The
beams ${\bf F}_{a_1}^{(+)}$ and ${\bf F}_{b_1}^{(+)}$ can be locally
manipulated, allowing for the possibility of distributing the vacuum
amplitudes in sixteen different ways, each corresponding to the
generation of a concrete Bell base state. Hence, the possibility of
performing superdense coding is explained in the WRHP framework through
the change of the correlation properties of the light beams
(represented by the four nonvanishing correlations $p
\leftrightarrow q$ and $r \leftrightarrow s$) when the two uncorrelated
beams corresponding to one photon are modified via local manipulations.
Such correlations have their origin in the crystal, where the zeropoint
modes are coupled with the laser field, and the information is carried
by the amplified vacuum fluctuations.

Now let us characterize the correlation properties of the sixteen Bell
base states. For this purpose the description of the four polarization
Bell-states, in terms of two-parametrized two correlated beams, will be
considered here \cite{pdc8}. For the sake of simplicity, from now on in
this section we shall discard the dependence on position and time. We
have:

\begin{equation}
{\bf F}_{x_1}^{(+)}=\left\{[F_s^{(+)}{\rm cos}\beta-F_p^{(+)}{\rm sin}\beta]
{\bf i}_{x_1}+{\rm e}^{i\kappa}[F_s^{(+)}{\rm sin}\beta+F_p^{(+)}{\rm cos}\beta]
{\bf j}_{x_1}\right\}{\rm e}^{i\varphi_1},
\label{hyper18a}
\end{equation}

\begin{equation}
{\bf F}_{y_1}^{(+)}=\left\{[{F'}_s^{(+)}{\rm cos}\beta-{F'}_p^{(+)}{\rm sin}\beta]
{\bf i}_{y_1}+{\rm e}^{i\kappa}[{F'}_s^{(+)}{\rm sin}\beta+{F'}_p^{(+)}{\rm cos}\beta]
{\bf j}_{y_1}\right\}{\rm e}^{i\varphi_2},
\label{hyper20a}
\end{equation}

\begin{equation}
{\bf F}_{a_2}^{(+)}={F'}_{q}^{(+)}{\bf i}_{a_2}+{F'}_{r}^{(+)}{\bf
j}_{a_2},
\label{hyper21a}
\end{equation}

\begin{equation}
{\bf F}_{b_2}^{(+)}=F_{q}^{(+)}{\bf i}_{b_2}+F_{r}^{(+)}{\bf
j}_{b_2},
\label{hyper19a}
\end{equation}
where ${\bf F}_{x_1}^{(+)}$ (${\bf F}_{y_1}^{(+)}$) is correlated to
${\bf F}_{b_2}^{(+)}$ (${\bf F}_{a_2}^{(+)}$).

Eqs. (\ref{hyper18a}) to (\ref{hyper19a}) correspond to the
description, in the WRHP, of the sixteen Bell base states corresponding
to polarization-momentum hyperentanglement of two photons. The
essential point is that quantum correlations are described in terms of
four two-by-two correlated beams through the eight sets of independent
zeropoint amplitudes entering the nonlinear source.

The transformations concerning polarization are represented by two
parameters, $\beta$ and $\kappa$, which represent the action of a
polarization rotator and a wave retarder, respectively, on beams
corresponding to photon ``$1$". In this way, the combination $\beta=0$,
$\kappa=0$ ($\beta=0$, $\kappa=\pi$) corresponds to the description of
the polarization state $|\Psi^+\rangle$ ($|\Psi^-\rangle$). In both
cases the non-null correlations correspond to different polarization
components, the only difference being the minus sign that appears in
$|\Psi^-\rangle$. On the other hand, the case $\beta=\mp\pi/2$ and
$\kappa=\pi$ ($\beta=-\pi/2$, $\kappa=0$) corresponds to the
description of $\mp|\Phi^+\rangle$ ($|\Phi^-\rangle$), where the
horizontal (vertical) component of a beam is correlated with the
horizontal (vertical) component of the conjugated one \cite{pdc8}.

On the other hand, momentum is represented by two couples of
parameters, $(x, y)$ and $(\varphi_1, \varphi_2)$, such that:

\begin{itemize}
\item The combination $(x, y)=(a, b)$ ($(x, y)=(b, a)$) and $\varphi_1=\varphi_2=0$ corresponds to the state
$|\psi^+\rangle$ ($|\phi^+\rangle$).

\item The situation in which $(x, y)=(a, b)$,
$\varphi_1=0$ and $\varphi_2=\pi$, corresponds to $|\psi^-\rangle$.

\item Finally, in the case $(x, y)=(b, a)$, $\varphi_1=\pi$ and $\varphi_2=0$,
we have the description of $|\phi^-\rangle$.
\end{itemize}

Because of the fact that multiplying a given Bell state by a phase
factor is irrelevant, it can be easily seen that there are four
dichotomic parameters giving rise to the sixteen Bell base states. For
instance, the parameters $\beta=(0, -\pi/2)$ and $\kappa=(0,
\pi)$ define the polarization Bell-state, and $[x, y]=[(a, b), (b,
a)]$ and $[\varphi_1, \varphi_2]=[(0, 0), (0, \pi)]$ the momentum
Bell-state. These parameters can be changed locally, and this property
is of interest in connection to dense coding and superdense coding
\cite{Walborn, 2007}.


\section{COMPLETE BSM IN THE WRHP}
\label{sec3}

In this section we shall apply the Wigner formalism for
hyperentanglement to the description of a complete BSM of the four Bell
states, by considering that one of the two degrees of freedom is in a
fixed state: ({\rm i})\,First, we shall consider the experimental setup
shown in Fig. 2 of reference \cite{Walborn}, in which the momentum
degrees of freedom are used as the ancilla, in order to encode
information in polarization Bell states; ({\rm ii})\,in the second
experiment (Fig. 3 of \cite{Walborn}), the polarization state is fixed,
and the four momentum Bell-states can be distinguished.

Both setups correspond to a broad class of LELM devices, in which the
modes corresponding to the input photons are not brought together in
the apparatus. In this case, it has been demonstrated that there are at
most $2^n$ distinguishable classes of Bell states \cite{2011}. As we
are considering experiments in which one of the degrees of freedom is
in a fixed state, each of the $2^2=4$ distinguishable classes of Bell
states will just correspond to one of the four Bell states of the other
degree of freedom.

In order to focus on the role of the zeropoint field, we shall describe
the different steps of the experiments. The values of the field
amplitudes at the detectors are usually computed by propagating them
through the optical devices from the source to the detectors. In these
experiments an identical distance separating the source from the
respective optical devices and detectors will be considered, so that
the contribution of the related phase shift in Eq. (16) of \cite{pdc4}
will be discarded in the calculation of the probabilities. For
simplicity, we shall focus on the ideal situation $t=t'$, so that we
can discard the dependence on position and time.

\subsection{Discrimination of the polarization Bell-states}

The hyperentangled-Bell-state analyser includes two polarizing
beam-splitters (PBS) which transmit (reflect) the vertical (horizontal)
polarization, and switch modes (remain in the same mode), in order to
perform a controlled-NOT (CNOT) logic operation between the
polarization (control) and spatial (target) degrees of freedom. Each
outgoing beam passes through a polarization analyzer (PA) in the $\pm
45^{\circ}$ basis, consisting of a half-wave plate, a PBS, and two
detectors. The vacuum zeropoint field at the idle channels of the
analyzers is represented in Fig. 2, this being an important ingredient
of our approach, as we shall see later.

\begin{figure}[h]
      \centering
      \includegraphics[height=10cm]{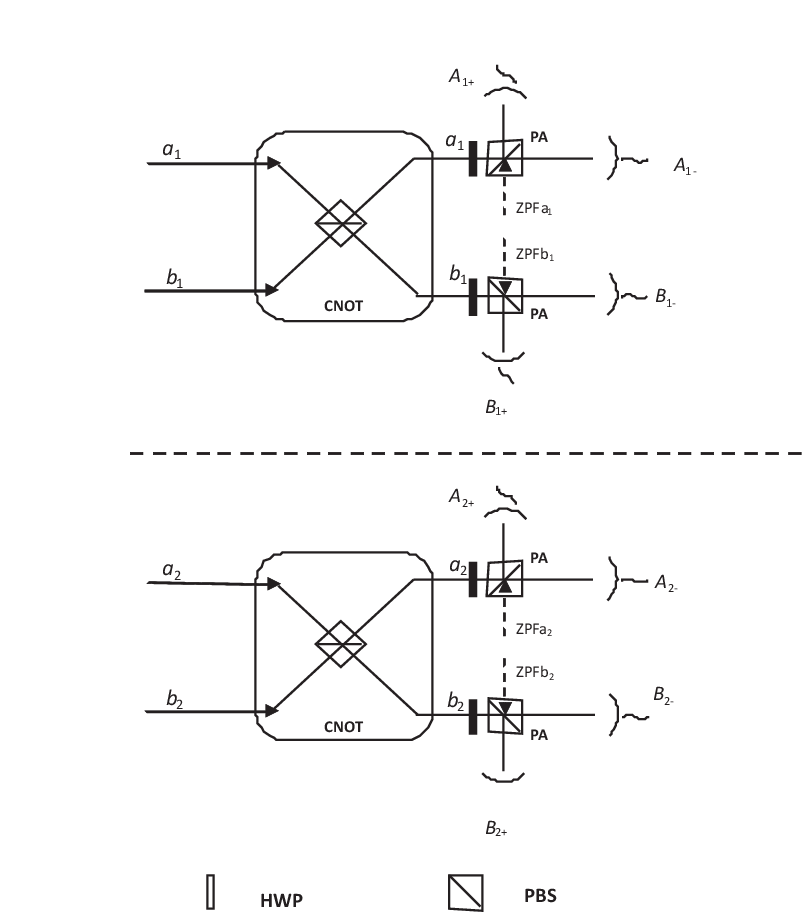}
 \begin{center}
\caption{Polarization-momentum hyperentanglement analyzer using a fixed entangled state in momentum.
The consideration of the zeropoint field at the idle channels of the
analyzers is a key point in the Wigner approach.}
\end{center}
\label{fig2}
\end{figure}

By considering $(x, y)=(a, b)$ and $\varphi_1=\varphi_2=0$ in Eqs.
(\ref{hyper18a}) to (\ref{hyper19a}), we obtain four correlated beams
which describe one of the four states: $|\Psi^\pm\rangle \otimes
|\psi^+\rangle$ and $|\Phi^\pm\rangle
\otimes |\psi^+\rangle$, depending on the value of $\beta$ and $\kappa$.
The action of CNOT gates is represented by the following output beams:
 \begin{equation}
{\bf {F'}}_{a_1}^{(+)}=i[F_s^{(+)}{\rm cos}\beta-F_p^{(+)}{\rm sin}\beta]
{\bf {i}}_{a_1}+{\rm e}^{i\kappa}[{F'}_s^{(+)}{\rm sin}\beta+{F'}_p^{(+)}{\rm cos}\beta]
{\bf {j}}_{a_1},
\label{hyper18aa}
\end{equation}
\begin{equation}
{\bf {F'}}_{b_1}^{(+)}=i[{F'}_s^{(+)}{\rm cos}\beta-{F'}_p^{(+)}{\rm sin}\beta]
{\bf {i}}_{b_1}+{\rm e}^{i\kappa}[F_s^{(+)}{\rm sin}\beta+F_p^{(+)}{\rm cos}\beta]
{\bf {j}}_{b_1},
\label{hyper18aaa}
\end{equation}
\begin{equation}
{\bf {F'}}_{a_2}^{(+)}=i{F'}_{q}^{(+)}{\bf {i}}_{a_2}+F_{r}^{(+)}{\bf
{j}}_{a_2},
\label{hyper21aaa}
\end{equation}

\begin{equation}
{\bf {F'}}_{b_2}^{(+)}=i
F_{q}^{(+)}{\bf i}_{b_2}+{F'}_{r}^{(+)}{\bf j}_{b_2},
\label{hyper19aaaa}
\end{equation}
where we have considered the imaginary unit in order to account for the
reflection at the PBS, in contrast to the equation (3) of reference
\cite{Walborn}, in which there is no phase shift associated to the
reflection.

A quick look at Eqs. (\ref{hyper18aa}) to (\ref{hyper19aaaa}) shows
that, in the case $\beta=0$, the nonzero correlations are those
concerning amplitudes related to orthogonal polarizations of beams
$a_1$ and $a_2$, or $b_1$ and $b_2$. This is consistent with the
transformation $|\Psi^{\pm}\rangle |\psi^+\rangle\rightarrow |\Psi^{\pm}\rangle
|\phi^+\rangle$ (see Eq. (4) of \cite{Walborn}). In contrast, in the
case $\beta=\pm \pi/2$ there is no change in the correlation properties
of the light beams. These properties justify that the momentum state
$|\psi^{+}\rangle$ can be used to discriminate between the four
polarization Bell states. It is noteworthy that there is no additional
zeropoint amplitudes entering the PBSs at the CNOT gates (see Fig. 2),
so that the PBSs mark the momentum state due to the consideration of
the eight sets of independent zeropoint modes at the two-crystal
source, and their subsequent redistribution in the beams' amplitudes.





Now, taking into account the action of the half-wave plate
-HWP$@45^{\circ}$-, and that the polarization analyzers are oriented at
$45^{\circ}$, the polarizing beam-splitters will reflect (transmit) the
component of the field along the unit vector ${\bf{i}}$ (${\bf j}$),
which is oriented at $+45^{\circ}$ ($-45^{\circ}$) with respect to the
horizontal direction. In order to express the field amplitudes at the
detectors, we must add the corresponding zero-point component that
enters through the free channel of each PBS. After some easy algebra,
we obtain the following compact expression for the four field
amplitudes at the detectors, concerning the paths $a_{1}$ and $b_{1}$:

\[
{\bf F}_{X_{1\pm}}^{(+)}
=
\frac{i^{n_{1\pm}}}{\sqrt{2}}\left\{i[\widetilde{F}_{s, X}^{(+)}
{\rm cos}\beta-\widetilde{F}_{p, X}^{(+)}{\rm
sin}\beta]\pm{\rm e}^{i\kappa}[\check{F}_{s, X}^{(+)}{\rm sin}\beta
+\check{F}_{p, X}^{(+)}\cos \beta ]\right\}{\bf{u}}_{\pm}
\]
\begin{equation}
+i^{m_{1\pm}}[{\bf{F}}_{ZPFX_{1}}^{(+)}
\cdot {\bf{u}}_{\mp}]{\bf{u}}_{\mp},
\label{compactw1}
\end{equation}
and for the paths $a_{2}$ and $b_{2}$

\begin{equation}
{\bf{F}}_{Y_{2\pm}}^{(+)}
=\frac{i^{n_{2\pm}}}{\sqrt{2}}[i\check{F}_{q, Y}^{(+)}
\pm \widetilde{F}_{r, Y}^{(+)}]
{\bf{u}}_{\pm}+i^{m_{2\pm}}[{\bf{F}}_{ZPFY_{2}}^{(+)}\cdot {\bf{u}}_{\mp}]{\bf{u}}_{\mp},
\label{compactw2}
\end{equation}
where: $X, Y=A$ or $B$; ${\bf{u}}_{+}\equiv \bf{i}$,
${\bf{u}}_{-}\equiv
\bf{j}$, and $n_{i+}=1$, $n_{i-}=0$, $m_{i+}=0$, $m_{i-}=1$, for $i=1,
2$. Also we have defined the amplitudes $\widetilde{F}_{v, Z}^{(+)}$
and $\check{F}_{v, Z}^{(+)}$ ($v=p, q, r, s$; $Z=X, Y$), where
$\widetilde{F}_{v, Z}^{(+)}=F_{v}^{(+)}$ in the case $Z=A$, and
$\widetilde{F}_{v, Z}^{(+)}={F'}_{v}^{(+)}$ stands for $Z=B$. In
contrast, $\check{F}_{v, Z}^{(+)}={F'}_{v}^{(+)}$ in the case $Z=B$,
and $\check{F}_{v, Z}^{(+)}={F}_{v}^{(+)}$ stands for $Z=A$.





In order to calculate the joint detection probabilities we shall use
Eq. (\ref{p12}) along with the correlation properties given in Eq.
(\ref{hyper16}). We shall take into account that the ZPF inputs at the
PBSs are uncorrelated with the signals and with each other. After some
easy calculations, we obtain the following general expression for the
joint detection probability:

\[
\frac{P_{X_{1\pm},Y_{2\pm}}}{k_{X_{1\pm}}k_{Y_{2\pm}}}
=\frac{1}{4}\left| [-i^2\langle\check{F}_{q, Y}^{(+)}\widetilde{F}_{p,
X}^{(+)}\rangle+(\pm)_1(\pm)_2{\rm e}^{i\kappa}\langle
\widetilde{F}_{r, Y}^{(+)}\check{F}_{s, X}^{(+)}\rangle]
{\rm sin}\beta\right.
\]
\begin{equation}
\left.+i [(\pm)_2\langle
\widetilde{F}_{r, Y}^{(+)}\widetilde{F}_{s, X}^{(+)}\rangle+(\pm)_1\langle\check{F}_{q, Y}^{(+)}\check{F}_{p,
X}^{(+)}\rangle]{\rm cos}\beta
\right|^2,
\label{compactprobw1}
\end{equation}
where $k_{X_{1\pm}}$, $k_{Y_{2\pm}}$, are constants which are related
to the effective efficiency of the detection process. Let us now
consider the following cases:

\begin{itemize}

\item {\it Case I} ($\beta=0$) corresponds to the states $|\Psi^\pm\rangle \otimes
|\psi^+\rangle$. From (\ref{compactprobw1}) it can be easily shown that
for $\kappa=0$ ($\kappa=\pi$), which corresponds to the polarization
state $|\Psi^{+}\rangle$ ($|\Psi^{-}\rangle$), only the four
probabilities $P_{A_{1+}, A_{2+}}$, $P_{B_{1+}, B_{2+}}$,
$P_{A_{1-},A_{2-}}$, $P_{B_{1-}, B_{2-}}$ ($P_{A_{1+}, A_{2-}}$,
$P_{B_{1+}, B_{2-}}$, $P_{A_{1-},A_{2+}}$, $P_{B_{1-}, B_{2+}}$) are
different from zero.

\item {\it Case II} ($\beta=\pm\pi/2$) corresponds to the states $|\Phi^\pm\rangle \otimes
|\psi^+\rangle$. In this case, from (\ref{compactprobw1}) it is shown
that for $\kappa=0$ ($\kappa=\pi$), which corresponds to the
polarization state $|\Phi^{-}\rangle$ ($|\Phi^{+}\rangle$), only the
four probabilities $P_{A_{1+}, B_{2+}}$, $P_{B_{1+}, A_{2+}}$,
$P_{A_{1-},B_{2-}}$, $P_{B_{1-}, A_{2-}}$ ($P_{A_{1+}, B_{2-}}$,
$P_{B_{1+}, A_{2-}}$, $P_{A_{1-}, B_{2+}}$, $P_{B_{1-}, A_{2+}}$) are
nonzero.

\end{itemize}

Hence, the detector signatures allow to distinguish between the
polarization Bell states. The discrepancy with Walborn's Table I in
ref. \cite{Walborn}, with respect to the states $|\Phi^{-}\rangle$ and
$|\Phi^{+}\rangle$, is due to the consideration of the complex factor
$i$ for the reflected amplitudes at the CNOT gates \cite{Zeilinger}.
For this reason, the corresponding joint probabilities for these two
states are exchanged with respect to the work of Walborn et al.

\subsection{Discrimination of the momentum Bell-states}

The experimental setup is shown in Fig. 3. Two half-wave plates (HWP),
which are aligned at $45^{\circ}$ in modes $b_1$ and $b_2$, perform the
CNOT operation. The BSs are balanced nonpolarizing beam splitters
\cite{Walborn}. In this case, the polarization degrees of freedom are
used as ancilla, so that the polarization state is fixed, and
corresponds to $|\Psi^{+}\rangle$. In the Wigner formalism, by putting
$\kappa=\beta=0$ in Eqs. (\ref{hyper18a}) to (\ref{hyper19a}), we
obtain the following four beams in order to compactly describe the four
states $|\Psi^+\rangle
\otimes |\psi^\pm\rangle$
and $|\Psi^+\rangle
\otimes |\phi^\pm\rangle$:

\begin{figure}
      \centering
      \includegraphics[height=10cm]{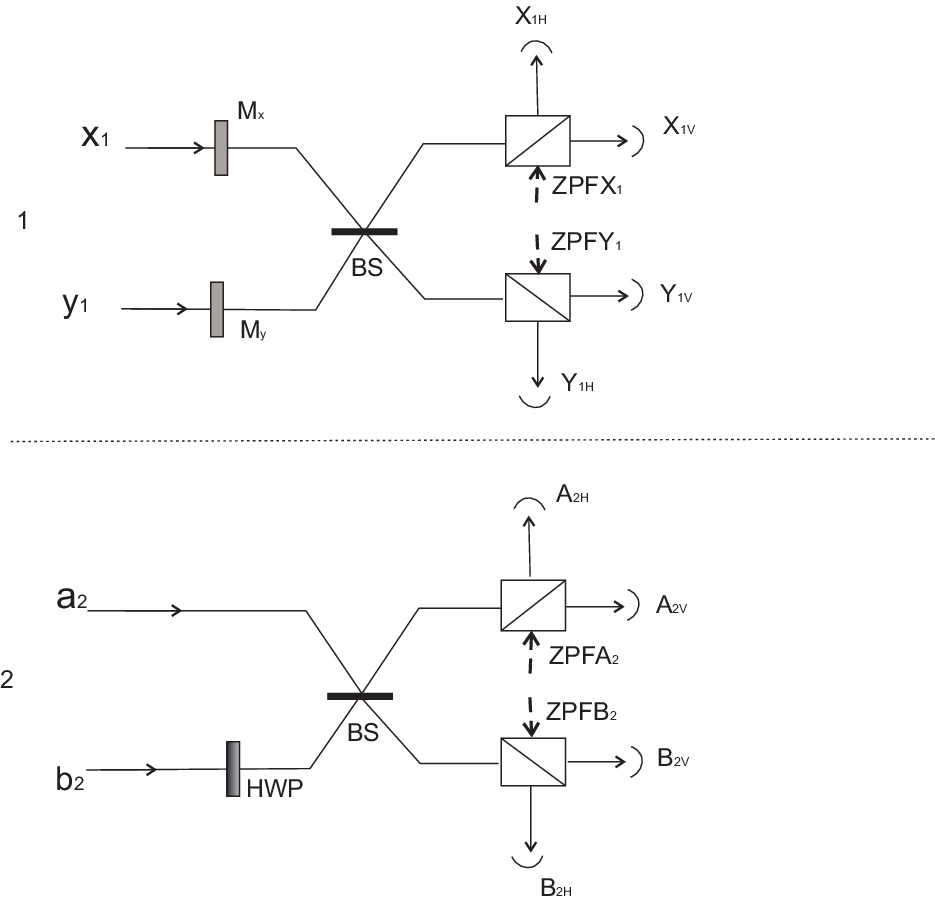}
 \begin{center}
\caption{Hyperentangled-Bell-state analyzer using a fixed entangled state in polarization.
We have represented two optical devices, $M_x$ and $M_y$, in order to
account for the two possibilities, depending on the value of $x$ and
$y$. The zeropoint field at the idle channels of the analyzers is
represented by four beams.}
\end{center}
\label{fig3}
\end{figure}

\begin{equation}
{\bf F}_{x_1}^{(+)}=[F_s^{(+)}
{\bf i}_{x_1}+F_p^{(+)}
{\bf j}_{x_1}]{\rm e}^{i\varphi_1},
\label{hyper18aaab}
\end{equation}

\begin{equation}
{\bf F}_{y_1}^{(+)}=[{F'}_s^{(+)}
{\bf i}_{y_1}+{F'}_p^{(+)}
{\bf j}_{y_1}]{\rm e}^{i\varphi_2},
\label{hyper20aaab}
\end{equation}

\begin{equation}
{\bf F}_{a_2}^{(+)}={F'}_{q}^{(+)}{\bf i}_{a_2}+{F'}_{r}^{(+)}{\bf
j}_{a_2}
\label{hyper21aaab}
\end{equation}

\begin{equation}
{\bf F}_{b_2}^{(+)}=F_{q}^{(+)}{\bf i}_{b_2}+F_{r}^{(+)}{\bf
j}_{b_2}.
\label{hyper19aaab}
\end{equation}

Now, we define the matrices $\hat{M_x}$ and $\hat{M_y}$: in the case
$x=a$, $y=b$, $\hat{M_x}=\hat{I}$ (identity matrix) and
$\hat{M_y}=\hat{M}_{HWP}$; on the other hand, if $x=b$, $y=a$, then
$\hat{M_x}=\hat{M}_{HWP}$ and $\hat{M_y}=\hat{I}$. In this way, the
beams entering the BSs are

\begin{equation}
{\bf{F'}}_{x_{1}}^{(+)}=\hat{M_x}{\bf{F}}_{x_{1}}^{(+)}\,\,\,\,;\,\,\,\,
{\bf{F'}}_{y_{1}}^{(+)}=\hat{M_y}{\bf{F}}_{y_{1}}^{(+)},
\label{joseba1}
\end{equation}



\begin{equation}
{\bf {F'}}_{a_2}^{(+)}={\bf {F}}_{a_2}^{(+)}\,\,\,\,;\,\,\,\,{\bf {F'}}_{b_2}^{(+)}=\hat{M}_{HWP}{\bf
{F}}_{b_2}^{(+)}= F_{r}^{(+)}{\bf i}_{b_2}+F_{q}^{(+)}{\bf j}_{b_2}.
\label{joseba2}
\end{equation}
From Eqs. (\ref{joseba1}) and (\ref{joseba2}) it can be easily seen
that, for $x=a$, $y=b$ ($x=b$, $y=a$), the only non-null
cross-correlations are those concerning the same polarization
(orthogonal polarizations). For this reason, the CNOT operation marks
the polarization state. This operation does not introduce additional
zeropoint fluctuations, because the HWPs do not activate zeropoint
modes. Finally, the BSs transform the beams given in Eqs.
(\ref{joseba1}) and (\ref{joseba2}) without the consideration of
additional zeropoint modes, because there is no idle channel at the
BSs.


Finally, if we consider the zeropoint amplitudes at the idle channels
of the PBSs (see Fig. 3), the field amplitudes at the detectors
$X_{1H}$, $X_{1V}$, $Y_{1H}$ and $Y_{1V}$ can be expressed by the
following compact expression:

\begin{equation}
F_{Z_{1\lambda}}^{(+)}
=
\frac{i^{n_{\lambda}}}{\sqrt{2}}\left\{i^{m_1(Z)}[\hat{M_x}{\bf{F}}_{x_{1}}^{(+)}]+
i^{m_2(Z)}[\hat{M_y}{\bf{F}}_{y_{1}}^{(+)}]\right\}\cdot {\bf
u}_{\lambda}+F_{vac, Z_{1\lambda}}^{(+)},
\label{porfin1}
\end{equation}
where: $Z=X, Y$, $\lambda=H, V$; ${\bf{u}}_{H}\equiv \bf{i}$,
${\bf{u}}_{V}\equiv
\bf{j}$; $n_{H}=1$, $n_{V}=0$; $m_1(X)=1$, $m_1(Y)=0$; $m_2(X)=0$, $m_2(Y)=1$.

In the same way, we have the following expression for the field
amplitudes at the detectors $A_{2H}$, $A_{2V}$, $B_{2H}$ and $B_{2V}$:

\begin{equation}
F_{V_{2\lambda'}}^{(+)}
=
\frac{i^{n_{\lambda'}}}{\sqrt{2}}\left\{i^{m_1(V)}{\bf{F}}_{a_{2}}^{(+)}+
i^{m_2(V)}[\hat{M}_{HWP}{\bf{F}}_{b_{2}}^{(+)}]\right\}\cdot {\bf
u}_{\lambda'}+F_{vac, V_{2\lambda'}}^{(+)},
\label{porfin2}
\end{equation}
where: $V=A, B$, $\lambda'=H, V$; $m_1(A)=1$, $m_1(B)=0$; $m_2(A)=0$,
$m_2(B)=1$.



Now, using Eqs. (\ref{hyper16}) and (\ref{p12}), and taking into
account that the ZPF inputs at the PBSs are uncorrelated with the
signals and with each other, we shall consider the following cases:

\begin{itemize}
\item {\it Case I} ($x=a$, $y=b$) corresponds to the states $|\Psi^+\rangle
\otimes |\psi^\pm\rangle$. After some easy algebra, we obtain:

\begin{equation}\begin{array}{l}\frac{P_{A_{1H},A_{2H}}}{k_{A_{1H}}k_{A_{2H}}}
=\frac{P_{B_{1H},B_{2H}}}{k_{B_{1H}}k_{B_{2H}}}=\frac{P_{A_{1V},A_{2V}}}{k_{A_{1V}}k_{A_{2V}}}
=\frac{P_{B_{1V},B_{2V}}}{k_{B_{1V}}k_{B_{2V}}}=
\frac{g^2|V|^2|\nu(0)|^2}{16}\left|e^{i\varphi_1}+e^{i\varphi_2}\right|^{2},
\label{sota9}
\end{array}\end{equation}
and

\begin{equation}\begin{array}{l}\frac{P_{A_{1H},B_{2H}}}{k_{A_{1H}}k_{B_{2H}}}
=\frac{P_{B_{1H},A_{2H}}}{k_{B_{1H}}k_{A_{2H}}}=\frac{P_{A_{1V},B_{2V}}}{k_{A_{1V}}k_{B_{2V}}}
=\frac{P_{B_{1V},A_{2V}}}{k_{B_{1V}}k_{A_{2V}}}=
\frac{g^2|V|^2|\nu(0)|^2}{16}\left|e^{i\varphi_1}-e^{i\varphi_2}\right|^{2}.
\label{sota10}
\end{array}\end{equation}

From Eqs. (\ref{sota9}) and (\ref{sota10}) it can be seen that, in the
case $\varphi_1=\varphi_2=0$ ($\varphi_1=0$, $\varphi_2=\pi$), which
corresponds to the momentum state $|\psi^+\rangle$ ($|\psi^-\rangle$),
only the four probabilities in Eq. (\ref{sota9}) (Eq. (\ref{sota10}))
are nonzero.


\item {\it Case II} ($x=b$, $y=a$) corresponds to the states $|\Psi^+\rangle
\otimes |\phi^\pm\rangle$. In this case, we obtain:

\begin{equation}\begin{array}{l}
\frac{P_{A_{1H},A_{2V}}}{k_{A_{1H}}k_{A_{2V}}}=\frac{P_{A_{1V},A_{2H}}}{k_{A_{1V}}k_{A_{2H}}}
=\frac{P_{B_{1H},B_{2V}}}{k_{B_{1H}}k_{B_{2V}}}=\frac{P_{B_{1V},B_{2H}}}{k_{B_{1V}}k_{B_{2H}}}=
\frac{g^2|V|^2|\nu(0)|^2}{16}\left|e^{i\varphi_1}-e^{i\varphi_2}\right|^{2}.
\label{sota11}
\end{array}\end{equation}




The above probabilities are non-null only in the case $\varphi_1=\pi$,
$\varphi_2=0$ (or viceversa), i.e. the WRHP description of the momentum
state $|\phi^-\rangle$.
On the other hand:

\begin{equation}\begin{array}{l}
\frac{P_{A_{1H},B_{2V}}}{k_{A_{1H}}k_{B_{2V}}}=\frac{P_{A_{1V},B_{2H}}}{k_{A_{1V}}k_{B_{2H}}}
=\frac{P_{B_{1H},A_{2V}}}{k_{B_{1H}}k_{A_{2V}}}=\frac{P_{B_{1V},A_{2H}}}{k_{B_{1V}}k_{A_{2H}}}
=\frac{g^2|V|^2|\nu(0)|^2}{16}\left|e^{i\varphi_1}+e^{i\varphi_2}\right|^{2}.
\label{sota13}
\end{array}\end{equation}


From Eq. (\ref{sota13}), the corresponding probabilities are non-null
only in the case $\varphi_1=\varphi_2=0$, i.e. the situation
corresponding to the momentum state $|\phi^+\rangle$.


\end{itemize}

There is a discrepancy with Walborn's Table II in ref. \cite{Walborn},
with respect to the states $|\phi^{-}\rangle$ and $|\phi^{+}\rangle$,
which is due to the consideration of the complex factor $i$ for the
reflected amplitudes at the BSs \cite{Zeilinger}. Then, the
corresponding joint probabilities for these two states are exchanged
with respect to the work of Walborn et al.



\section{ZPF AND THE LIMITS ON OPTIMAL BSM}
\label{sec4}

Let us consider the following situation: two photons entangled in $n$
dichotomic degrees of freedom enter a LELM apparatus via separate
spatial channels, designated $L$ and $R$. Each photon contains $2^n$
modes, and a unitary matrix transforms the $2^{n+1}$ input modes to
$2^{n+1}$ output modes to the detectors. In \cite{2011} it is
demonstrated that: (i)\,A single detector click cannot discriminate
between any of the Bell states, so that the $2^{n+1}$ possibilities for
the second detector event form a simple upper bound on distinguishable
Bell-state classes using LELM devices; (ii) There can be at most
$2^{n+1}-1$ distinguishable classes of hyper-Bell states for two
bosons. This number is reduced to $2^n$ in the case that the left and
right channels are not brought together in the apparatus. The
experiments described in Sec. \ref{sec3} are just an example of this
situation in the case of photons, for $n=2$. This section is divided in
two parts: (\ref{en1})\,First, we shall investigate the relationship
between the statement (i) and the zeropoint field inputs when optical
experiments using parametric down conversion are considered;
(\ref{en2})\, We shall study the relationship between the number of
vacuum inputs at the source and the analysers, and the maximum number
of distinguishable classes, $2^n$, in the kind of experiments as the
ones described in Sec. \ref{sec3}.

\begin{enumerate}

\item \label{en1} Given an optical n-qubit state, the maximum number of mutually
distinguishable sets of Bell states is bounded above by $2^{n+1}$
\cite{2007, 2011}. The demonstration of this point in the Hilbert space
is based on a particle-like description, which contrasts to the image
in the WRHP formalism. For instance, in \cite{pdc8} it was stressed
that two-photon entanglement in only one degree of freedom implies the
consideration of four independent sets of zeropoint modes at the
source, which are ``activated" through a coupling with the laser inside
the crystal. In this paper, the generation of polarization-momentum
hyperentanglement is represented via the consideration of eight sets of
independent vacuum modes, which are amplified at the two-crystal
source. Hence, hyperentanglement, i.e. entanglement in Hilbert spaces
of higher dimensions is closely related to the inclusion of more sets
of vacuum modes entering the source. With an increasing number of
vacuum inputs, the possibility of extracting more information from the
zeropoint field also increases. As we shall demonstrate below, for a
given $n$, the maximal distinguishability in a Bell-like experiment is
bounded by the number of independent vacuum sets of modes which are
extracted at the source, being this number equal to $2^{n+1}$. In order
to prove this statement, we shall consider the following lemmas:

{\it Lemma I:}
{\it For a two-photon n-qubit state generated via PDC, the number of
independent sets of zeropoint modes which are necessary for the
generation of entanglement, is just $2^{n+1}$.}

{\it Proof:} By considering the situation described in this paper, in which
$n=2$, the interaction Hamiltonian given in Eq. (\ref{free1}) gives
rise to linear evolution equations for the amplitudes $\alpha_{{\bf k},
\lambda}$ [see Eq. (\ref{eqb5})]. Hence, the $2^{n+1}$ sets of
field amplitudes outgoing the crystal are generated from an identical
number of independent sets of ZPF amplitudes entering the nonlinear
source, which are ``amplified" via the coupling with the laser beam.
This result is also true for $n=1$, as it has been demonstrated
elsewhere \cite{pdc4, pdc8}. For $n>2$, the interaction Hamiltonian is
also quadratic, which implies that the evolution equations for the
vacuum amplitudes are linear. Hence, this result is valid for any
$n\geq 1$.

{\it Lemma II:} {\it The propagation of the $2^{n+1}$ sets of field amplitudes outgoing the
crystal through a LELM device gives rise to $2^{n+1}$ output amplitudes
to the detectors. Each of them will include, at least, the $2^{n+1}$
sets of input ZPF amplitudes at the source.}

{\it Proof:} Given the fact that linear mode transformations lead to Bogoliubov
transformations of the mode operators, which are generated via
quadratic Hamiltonians \cite{kok}, the total set of unitary
transformations, including the generation of PDC light at the crystal
and the action of linear optical devices between the nonlinear source
and the detectors, are represented by quadratic Hamiltonians in mode
operators, which give rise to linear equations (see Fig. 4). When
passing to the Wigner representation, in which the destruction
(creation) operator $\hat{a}_{{\bf k}, \lambda}$
($\hat{a}^{\dagger}_{{\bf k},
\lambda}$) is substituted by a complex amplitude $\alpha_{{\bf k},
\lambda}$ ($\alpha_{{\bf k},
\lambda}^{*}$), each of the $2^{n+1}$ output field amplitudes at the detectors will include,
generally, the $2^{n+1}$ sets of field amplitudes outgoing the crystal,
and so the $2^{n+1}$ input ZPF amplitudes at the nonlinear source (see
Lemma I).

{\it Lemma III:}
{\it The $2^{n+1}$ possibilities for the second detection event, which
constitutes a simple upper bound on distinguishable Bell-state classes
from LELM devices, is just the number of the input ZPF sets of modes at
the source.}

{\it Proof:}
This result follows from lemmas $I$ and $II$.

Hence, the upper bound of $2^{n+1}$, which limits the optimality of any
Bell-state analysis, coincides with the number of relevant ZPF modes at
the source. This key point represents the relationship between the
information that can be effectively measured in an experiment and the
ZPF inputs at the source. In other words, the number of relevant sets
of zeropoint modes at the source represents a limit on optimal
Bell-state analysis in the enlarged Hilbert space.

From \cite{2011}, it is well known that a single detector event cannot
discriminate any of the Bell base states. Let us consider that a PBS is
placed at each of the beams (\ref{hyper18a}) to (\ref{hyper19a}) in
order to measure the single and joint detection probabilities,
following a setup similar to Fig. 1 of Ref. \cite{pdc7}, and taking
into account the vacuum inputs at the idle channels of the PBSs. Using
the autocorrelation properties of the light field given in Eq.
(\ref{eqb}), the expression (\ref{probsimple}) for the single detection
probability, and the field amplitudes at the detectors, it can be
easily demonstrated that the single detection probabilities are
identical, and independent of the parameters $\beta$, $\kappa$, $(x,
y)$ and $(\varphi_1,
\varphi_2)$.
The same reasoning holds for the experiments developed in
Sec.\ref{sec3}, by using the field amplitudes at the detectors given in
Eqs.  (\ref{compactw1}) and (\ref{compactw2}) [(\ref{porfin1}) and
(\ref{porfin2})] for the first (second) experiment.

\begin{figure}[h]
      \centering
      \includegraphics[height=5cm]{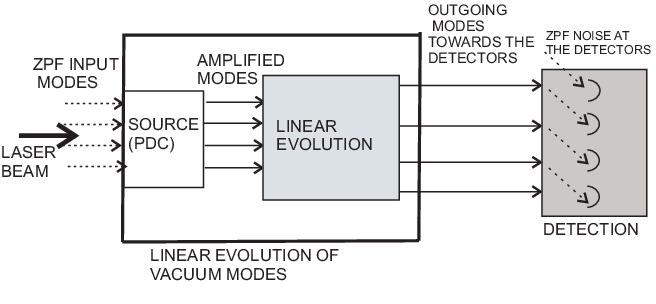}
 \begin{center}
\caption{The ZPF inputs at the non linear source are amplified and propagate through linear devices.
The total set of transformations are represented by a quadratic
Hamiltonian which gives rise to linear evolution equations. The total
number of modes outgoing towards the detectors is equal to the entering
ZPF modes at the crystal, which are amplified via the coupling with the
laser beam. We have also represented the ZPF inputs at the detectors. }
\end{center}
\label{fig4}
\end{figure}

\item
\label{en2} The noise entering the idle channels of the analyzers limits the
optimality of the Bell-state analysis, and this idea is worthy of
consideration. In Section \ref{sec3} we have applied the WRHP formalism
to complete BSM, in the case where one of the degrees of freedom is in
a fixed (ancillary) state, and the information is encoded in the other
degree of freedom. These experiments correspond to a general class of
setups in which the two photons are not mixed in the apparatus, so that
the maximun number of distinguishable classes of Bell states, is just
$2^n$ \cite{2011}. From the point of view of the WRHP approach, this is
exactly the number of non-vanishing cross-correlations between the
field amplitude at each detector on the left (right) side, and the
whole set of amplitudes at the detectors on the right (left) side, in
the optimal situation of maximal distinguishability.

On the other hand, each cross-correlation property of the field of the
kind of Eq. (\ref{hyper16}) is related to the probability of a joint
detection, in which the subtraction of the zeropoint intensity at each
detector is relevant (\ref{prob}). In order to be able to measure such
a correlation, the beam has to be divided, being necessary a zeropoint
contribution through the idle channel of the PBS, in order to preserve
the commutation relations. This zeropoint beam introduces two sets of
vacuum modes, one of them corresponding to vertical polarization and
the other to the horizontal one, which are uncorrelated with the signal
entering the other channel.

For instance, if $n=1$ ($n=2$), i.e. the photon pair is described by
two (four) correlated beams, and two (four) correlations are generated
by the source between the ``amplified" zeropoint fluctuations, almost
two (four) entry points of noise are necessary for measuring such
correlations, so that four (eight) sets of zeropoint amplitudes enter
the analyser. In the general case of $n$ degrees of freedom, the total
number of entry points of noise at the analysers is just $2^n$, so that
$2^{n+1}$ sets of vacuum amplitudes must be taken into account at the
Bell state analyser.

The net effect of the vacuum inputs at the BSM station is to decrease
the optimality of the Bell-state analysis. Following \cite{2011}, given
a detection at, for instance, the left side, the $2^{n}$ possible joint
detections at the right side, in the case in which the photons are not
mixed at the LELM device, gives the maximaum number of sets of Bell
states that can be distinghished in these experiments. However,
following the WRHP formalism, the upper bound, $2^{n+1}$, is given by
the number of ZPF sets of modes which are amplified at the source.
Which quantity must be subtracted from $2^{n+1}$ in order to obtain
$2^n$? The answer is just $2^n$, which constitutes the number of
zeropoint sets of vacuum modes on the right side (the area
corresponding to the second detection), and also the total number of
entry points of noise at the analyser. For instance, in Walborn's
experiments we can determine the number of Bell states classes that can
be distinguished by subtracting the number of channels with noise from
the number of ZPF entry modes, so that in this case we have $8-4=4$.
Due of the fact that one of the degrees of freedom is in a fixed state,
this number will coincide with the number of the Bell base states
corresponding to the other degree of freedom.

Hence, for a given $n$, if $N_{ZPF, S}=2^{n+1}$ is the number of sets
of zeropoint modes at the source, $N_{ZPF, x}=2^n$, $x=L, R$, is the
number of sets of vacuum modes entering the idle channels of the
analysers at the left or right area, and $N_{ic}=2^n$ is the number of
idle channels (entry points of noise) at the analyser, the maximum
number of mutually distinguishable classes of Bell states, $N_{max,
class}$, will be given by:

\begin{equation}
N_{max, class}=N_{ZPF, S}-N_{ZPF, x}=N_{ZPF, S}-N_{ic}.
\label{sisisi}
\end{equation}

\end{enumerate}

\section{CONCLUSIONS}
\label{sec5}
The zeropoint field at the optical experiments on quantum information
is not merely a mathematical tool which gives rise, after being
subtracted, to a broad class of theoretical results. From our point of
view, the vacuum field has a ``visible" presence in these experiments,
and this is what we are trying to demonstrate in this paper. Using the
WRHP approach we have analysed polarization-momentum hyperentanglement
in detail, showing the close relationship between enlarging Hilbert
spaces and ZPF inputs at the source, so that the possibility for
extracting more information, when an enlarged Hilbert space is used, is
due to the consideration of a greater number of ZPF inputs which are
amplified at the source.

Eqs. (\ref{hyper18a}) to (\ref{hyper19a}) give the WRHP description of
the sixteen Bell-like states in terms of four two-by-two correlated
beams. Each of the states is described by giving the value of four
parameters: two of them ($\beta$ and $\kappa$) indicate the
polarization Bell-state, and the other two, the dichotomic couples
($x$, $y$) and ($\varphi_1$, $\varphi_2$), are related to the momentum.
Let us emphasize that these parameters can be locally controlled
because all of them appear in Eqs. (\ref{hyper18a}) and
(\ref{hyper20a}), corresponding to photon $1$. Hence dense coding
\cite{Walborn} and superdense coding \cite{2007} are justified, in the
context of WRHP, by the possibility of changing the correlation
properties of the four beams through the action of a linear optical
device which operates in the same way as in classical optics. This
contrasts with the usual description in the Hilbert space, in which
local operations are represented by unitary operators. In the Wigner
framework, the effect of a linear optical device on a beam accounts for
a change in the distribution of the zeropoint amplitudes inside the
field components, so that there is a change in the correlation
properties. Given that these operations do not introduce additional
zeropoint noise, the information encoded in the zeropoint amplitudes
entering the source can be manipulated in order to succesfully complete
a quantum dense coding protocol.

We have analysed two experimental setups for complete BSM, each using a
fixed state in one of the two degrees of freedom, so that the
information is encoded in the other, as it appears in reference
\cite{Walborn}. As we have already pointed out, once within the Wigner
framework, the typical quantum results appear precisely as a
consequence of the role of the zeropoint field in the production,
propagation and detection of light. Quantum correlations can then be
explained solely in terms of the propagation of those vacuum amplitudes
through the experimental setup, and their subsequent subtraction at the
detectors. Hence, the Wigner formalism allows for an interpretation of
these experiments in terms of waves, where photons are just
wave-packets carrying the zeropoint amplitudes through the experimental
setup, and finally detected.


We have explained how the zeropoint inputs contribute to
distinguishability in a Bell-state analysis in which both
down-converted photons are not brought together at the LELM device. In
this situation, the difference between the amplified zeropoint modes at
the source ($2^{n+1}$) and the ZPF inputs at the Bell-state analyser
($2^n$), gives the maximal distinguishability of Bell-state classes
($2^n$). The influence of the zeropoint inputs in a general strategy of
Bell-state measurement, in which both photons are brought together at
the apparatus \cite{Vaidman}, will be the aim of further research.

The main conclusion of this work is that the WRHP approach allows for
the possibility of obtaining additional information to the one provided
by the standard Hilbert space formalism, just by considering the role
of the zeropoint field at the different steps of an optical quantum
communication experiment using PDC. Likewise, there is a close
relationship between the zeropoint extracted at the source, the
corresponding zeropoint field entering the vacuum channels of the
analyzers, and the maximal information that can be extracted in a
concrete experiment, as we have discussed in Sec.\ref{sec4}. This idea
will be developed in further works.

\section{ACKNOWLEDGEMENTS}
The authors would like to thank Prof. E. Santos for revising the
manuscript, and for helpful suggestions and comments on the work. A.
Casado acknowledges the support from the Spanish MCI Project no.
FIS2011-29400.

\appendix
\numberwithin{equation}{section}
\section{Appendix: General aspects of the WRHP}\label{Appendix}



The Wigner transformation establishes a correspondence between a field
operator acting on a vector in the Hilbert space and a (complex)
amplitude of the field. In the context of PDC the electric field
corresponding to a signal generated by the source (placed at ${\bf
r}=0$) is represented by a slowly varying amplitude \cite{pdc4}:

\begin{equation}
{\bf F}_{s}^{(+)}({\bf r}, t)=i{\rm e}^{\omega_s
t}\sum_{{\bf k}\in [{\bf k}]_{s}, \lambda=H, V}\left(\frac{\hbar
\omega_{{\bf k}}}
{2\epsilon_0L^3}\right)^{\frac{1}{2}}\alpha_{{\bf k}, \lambda}(t){\bf
u}_{{\bf k}, \lambda}{\rm e}^{i{\bf k}\cdot{\bf r}},
\label{F}
\end{equation}
where $[{\bf k}]_{s}$ represents a set of wave vectors centered at
${\bf k}_s$, and $\omega_s$ is the average frequency of the beam. ${\bf
u}_{{\bf k}, \lambda}$ is a unit polarization vector. In the Heisenberg
picture all the dynamics is contained at the amplitudes $\alpha_{{\bf
k},\lambda}(t)$, while the Wigner function is time-independent. In PDC,
the initial state is the vacuum, which is characterized by an electric
field given by (\ref{F}), by putting $\alpha_{{\bf k},
\lambda}(t)=\alpha_{{\bf k}, \lambda}{\rm exp}(-i\omega_{{\bf k}}t)$, where
$\alpha_{{\bf k}, \lambda}$ represents the zeropoint amplitude
corresponding to the mode $\{{\bf k}, \lambda\}$. The Wigner
distribution for the vacuum field amplitudes is \cite{pdc2}:

\begin{equation}
W_{\it ZPF}(\{\alpha\})=
{\prod_{{\bf [k]}, \lambda}}\frac{2}{\pi}
{\rm e}^{-2|\alpha_{{\bf k}, \lambda}|^2},
\label{eq_w9}
\end{equation}
where $\{\alpha\}$ represents the set of zeropoint amplitudes. Given
two complex amplitudes, $A({\bf r}, t; \{\alpha\})$ and $B({\bf r'},
t';\{\alpha\})$, the correlation between them is given by:

\begin{equation}
\langle AB \rangle \equiv \int W_{\it ZPF}(\{\alpha\})A({\bf r}, t; \{\alpha\})B({\bf r'}, t';
\{\alpha\})d\{\alpha\}.
\label{corr}
\end{equation}
For instance, from (\ref{eq_w9}) the well known correlation properties
hold:

\begin{equation}
\langle \alpha_{{\bf k}, \lambda}\alpha_{{\bf k'}, \lambda'}\rangle
=\langle \alpha^*_{{\bf k}, \lambda}\alpha^*_{{\bf k'},
\lambda'}\rangle=0\,\,\,\,;\,\,\,\,\,\langle \alpha_{{\bf k}, \lambda}\alpha^*_{{\bf k'}, \lambda'}\rangle
=\frac{1}{2}\delta_{{\bf k}, {\bf k'}}\delta_{\lambda, \lambda'}.
\label{correlations}
\end{equation}

In the Wigner approach, the single and joint detection probabilities in
PDC experiments are calculated by means of the expressions \cite{pdc4}:

\begin{equation}
P_{A}\propto \langle I_A-I_{ZPF, A} \rangle,
\label{probsimple}
\end{equation}

\begin{equation}
P_{AB}\propto
\langle(I_A-I_{ZPF, A})(I_B-I_{ZPF, B}) \rangle,
\label{prob}
\end{equation}
where $I_i\propto {\bf F}_{i}^{(+)}{\bf F}_{i}^{(-)}$, $i=A, B$, is the
intensity of light at the position of the i-detector, and $I_{ZPF, i}$
is the corresponding intensity of the zeropoint field. In experiments
involving polarization, the following simplified expression for the
joint detection probability will be used for practical matters:

\begin{equation}
P_{AB} \left({\bf{r}},t;{\bf{r'}},t'\right)\propto
\sum _{\lambda }\sum _{\lambda '}
\left|\left\langle F_{\lambda }^{\left(+\right)} \left(\phi_{A} ;{\bf{r}},t\right)
F_{\lambda '}^{\left(+\right)} \left(\phi_{B} ;{\bf{r'}},
t'\right)\right\rangle \right|^{2},
\label{p12}
\end{equation}
where and $\phi_{A}$ and $\phi_{B}$ are controllable parameters of the
experimental setup.

\end{document}